\documentclass[%
 reprint,
 amsmath,amssymb,
 aps,
]{revtex4-2}

\usepackage{graphicx}
\usepackage{dcolumn}
\usepackage{bm}
\usepackage{multirow}

\usepackage{xcolor}
\begin{document}

\preprint{APS/123-QED}

\title{Hydrogen segregation around \\a  straight screw dislocation in bcc iron}

\author{Margot LUCAS\textsuperscript{1,2}}
\author{Marie LANDEIRO DOS REIS\textsuperscript{1}}%
\author{Sylvain QUEYREAU\textsuperscript{2}}
\author{Xavier FEAUGAS\textsuperscript{1}}

\affiliation{%
 \textsuperscript{1} Laboratoire des Sciences de l’Ingénieur pour l’Environnement (LaSIE) - UMR CNRS 7356\\
}%
\affiliation{
 \textsuperscript{2} Laboratoire des Sciences des Procédés et des Matériaux (LSPM) - UPR 3407\\
}%

\date{May 4th 2026}

\begin{abstract}

The interaction between hydrogen and screw dislocations in bcc iron is central to understanding hydrogen embrittlement. A major challenge lies in the high-dimensional parametric landscape governing this interaction. In this work, we perform a comprehensive set of molecular simulations using a reliable neural network interatomic potential, systematically exploring hydrogen binding across dislocation core structures (easy and hard cores), site types, and concentrations. From these energetics, we construct a thermodynamic framework that quantifies the statistical relevance of the various trapping configurations, thereby significantly reducing the complexity of the problem. Our results show good agreement with the limited density functional theory data available in the literature. We further delineate the validity domain of an elastic dipole description of hydrogen–dislocation interactions, providing a simplified yet physically grounded modeling approach. Finally, we demonstrate that the easy-core configuration plays a key role in rationalizing experimental hydrogen solubility limits. These findings establish a consistent multiscale foundation for incorporating hydrogen–dislocation interactions into larger-scale models of plasticity and embrittlement.
\end{abstract}

\maketitle


\section{\label{level1}Introduction}

Hydrogen is ubiquitous in nature and is regarded as one of the solution to produce clean
energy to meet neutral carbon emission objectives in the future. A growing number of materials will thus be facing hydrogen and its presence often lead to important change of microstructure and properties, up to a loss of ductility known as hydrogen embrittlement (HE). This is particularly true for ferritic steels employed in gaz pipeline foreseen as a way to transport hydrogen. Several mechanisms have been proposed in the literature to explain the HE such as like Hydrogen Enhanced Decohesion \cite{ oriani1972, troiano2016, Ehlers2020} (HEDE) or enhanced localized plasticity \cite{beachem1972, birnbaum1994} (HELP) or Adsorption-Induced Dislocation Emission (AIDE) \cite{lynch2012, feaugas2023relationship} or H-enhanced strain-induced vacancies (HESIV) \cite{nagumo_book_2016} (see also reviews \cite{lynch2012, robertson2015, barrera2018, yu_chemrev_2024}). However, the wide range of hydrogen effects across multiple time and length scales, combined with the persistent difficulty in determining its exact localization within microstructures, makes it challenging to establish a unified description. It is therefore increasingly clear that only a comprehensive, multiscale framework can provide robust and predictive insights into HE.

Within this context, dislocations mediate most proposed hydrogen embrittlement mechanisms. In bcc iron, $a_0/2⟨111⟩$ screw dislocations govern plastic flow up to ~room temperature due to their non-planar, sixfold-symmetric core structure \cite{frederiksen2003, domain2005,dezerald2014ab}, which alternates between a stable easy-core and metastable hard-core configuration. The resulting Peierls barrier—stemming from lattice friction—exhibits complex stress and temperature dependence, as revealed by density functional theory (DFT) \cite{dezerald2014ab, clouet2021} and molecular dynamics (MD) simulations \cite{domain2005, gilbert2011, Gilbert:2013, rodney2009, freitas2018}.

When present, H atoms occupy tetrahedral sites in the bcc bulk, consistent with their low solubility. Near screw dislocations, octahedral sites become energetically favorable. Atomistic studies reveal a complex H interaction energy landscape around both easy-core (EC) and hard-core (HC) configurations \cite{itakura2012first, itakura2013effect, simpson2020effect, huang2023quantitative, zheng2023atomistic}, with H inducing a competition between core structures: EC offers more binding sites, while HC provides deeper wells. At high H concentrations, HC becomes favored, with thermodynamic models predicting H ordering and HC saturation. Yet measured H solubility in ferritic and pearlitic steels under gaseous loading remains low ($\approx 50–500$ appm upon deformation \cite{moro2009fragilisation}), and variations in H-dislocation interaction energy directly correlate with changes in the Peierls barrier for straight screw glide. Quantifying these interactions is thus essential to rationalize hydrogen's influence on screw dislocation mobility.

One key challenge lies in navigating the vast parametric hyperspace governing H-screw interactions, encompassing bulk and dislocation-core sites, H concentration, and easy-core (EC) vs. hard-core (HC) configurations.

Here, we propose a statistical physics framework to identify the dominant configurations, starting from a detailed mapping of H-Fe interactions via machine-learning (ML) interatomic potential \cite{meng2021general}. From this, we develop a thermodynamic canonical ensemble model incorporating EC/HC states and dislocation density. We further assess the viability of a cost-effective elastic dipole approximation \cite{Gillan_1984, nazarov2016first, clouet2018elastic} as a surrogate for explicit atomistic calculations, delineating its application limits—a critical input for mesoscale simulations such as kinetic Monte Carlo (KMC) and discrete dislocation dynamics (DDD). The impact of H on screw glide via the double-kink mechanism will be addressed in forthcoming work.


\section{\label{Methodology}Methodology}
Molecular statics simulations were carried out with the Large-scale Atomic/Molecular Massively Parallel Simulator (LAMMPS) \cite{thompson_lammps_2022}. The forces were computed using the Fe–H neural-network interatomic potential developed by Meng \textit{et al.} \cite{meng2021general}, which accurately reproduces DFT results for the key quantities governing hydrogen–dislocation interactions in bcc iron, including screw-dislocation core structures, the Peierls barrier and hydrogen solution energies- features that are not simultaneously captured by the classical empirical potentials tested in this work (see Tab.~\ref{tab:comparisonPot}).

\begin{table}[ht!]
\caption{\label{tab:comparisonPot}
Comparison of interatomic potentials NNIP\cite{meng2021general}, MEAM\cite{lee2007modified}, EAM\cite{wen2021new} with DFT results \cite{itakura2012first,borges2022ab,luthi2017modelisation,luthi2018attractive}. Relative core stability is evaluated via energy difference between EC and HC configurations $\Delta E_{HC-EC}$ and between split core (SC) and EC configurations $\Delta E_{SC-EC}$ (meV/$|\mathbf{b}|$). $a_0$ is the lattice parameter (\AA), $C_{11}$, $C_{22}$, $C_{44}$ are elastic constants (GPa). Peierls energy $E_P$ and H solution energy $E_{sol}$ are given in eV.}


\begin{ruledtabular}
\begin{tabular}{c|c|c|c|c}
 & \multicolumn{3}{c|}{This work} &  \multirow{2}{*} {DFT}\\
 & NNIP & MEAM  & EAM  &\\
\colrule

$\Delta E_{HC-EC}$&   40 & 50 &4 & 40 \cite{itakura2012first,luthi2017modelisation}, 80 \cite{luthi2018attractive}\\
$\Delta E_{SC-EC}$&  NP* & -50 &11 & 109 \cite{itakura2012first}\\
$a_0$  & 2.83 & 2.86& 2.86 & 2.83 \cite{itakura2012first,borges2022ab,luthi2017modelisation}\\
$C11$ & 296 & 243 & 243 & 237 \cite{itakura2012first}, 299 \cite{borges2022ab,luthi2017modelisation} \\
$C12$ & 147 & 138 & 144 & 104 \cite{itakura2012first}, 153 \cite{borges2022ab,luthi2017modelisation} \\
$C44$& 96 & 121 & 116 & 116 \cite{itakura2012first}, 103 \cite{borges2022ab,luthi2017modelisation} \\
$E_{P}$ & 38 & -50 & 11 & 40 \cite{itakura2012first}\\
$E_{sol}$ & 0.23 & 0.18 & 0.18 & 0.22 \cite{borges2022ab}\\
\end{tabular}
\end{ruledtabular}
\end{table}

Atomic relaxations were performed using the FIRE algorithm available in LAMMPS \cite{guenole2020assessment}, and all configurations were converged until the residual atomic forces fell below $10^{-3}$ eV/\AA. Atomic configurations were visualized and analyzed with OVITO \cite{stukowski_visualization_2009}.

\begin{figure}[ht!]
\includegraphics[width=8.5cm]{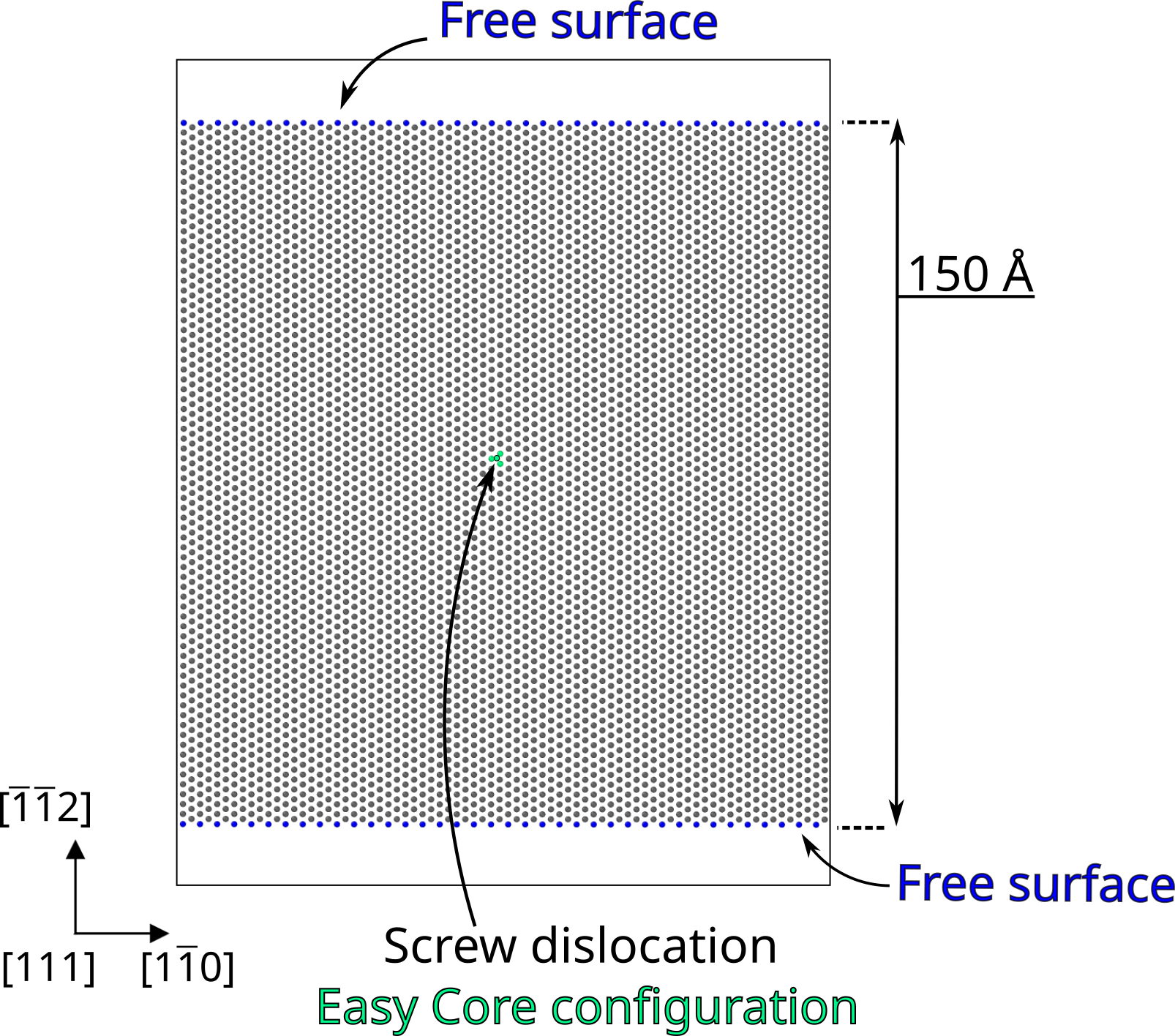}
\caption{\label{fig:system} Simulation cell with a screw dislocation (in green) at the center, in easy core configuration. We choose to have a free surface (in blue) following $[\overline{1}\overline{1}2]$ for periodic boundary condition (avoid artifact due to periodicity).}
\end{figure}

To compute the interaction energy between the dislocation core and hydrogen, the simulation cell was oriented along the crystallographic directions $x=[1\overline{1}0]$, $y=[\overline{1}\overline{1}2]$, and $z=[111]$. The $[111]$ direction aligns with the line direction $1/2\langle 111\rangle$ of screw dislocations in bcc iron, providing direct access to the $\left\{110\right\}$ and $\left\{112\right\}$ slip planes on which these dislocations predominantly glide.

Periodic boundary conditions were applied along the $y$ and $z$ directions, whereas free surfaces were imposed along $x$ (see Fig.~\ref{fig:system}). The system contained $N_{Fe}$ Fe atoms and a single hydrogen atom ($n_H = 1$). The simulation cell dimensions were set to 150 \AA{} along both $x$ and $y$ directions. Along $z$, a length of 3$b$ was used for the high-hydrogen-concentration case (approaching H saturation along the dislocation line), while a length of 10$b$ was adopted for the dilute limit. This larger cell size was selected to minimize image interactions and to ensure a reliable description of the dislocation–hydrogen interaction (see App. \ref{App:cellsize}).

The interaction energy is computed as:
\begin{equation}
E^{MS}_{\text{int}} = E_{D,H} - E_{D} - n_H\, \delta E_H ,
\label{eq:Eint}
\end{equation}
where $E_{D,H}$ is the energy of the cell containing both the dislocation and the hydrogen atom, $E_{D}$ is the energy of the cell containing only the screw dislocation, and $E_{H}$ denotes the energy cost of introducing a hydrogen atom into the cell. We define $\delta E_H = E_{H} - E_{0}$, where $E_{0}$ and $E_{H}$ correspond to the energies of the pristine cell and of the cell containing a single hydrogen atom, respectively.

\section{\label{Results}Results and discussion}
\subsection{Interaction energy}
\begin{figure}[ht!]
    \centering
    \includegraphics[width=0.6\linewidth]{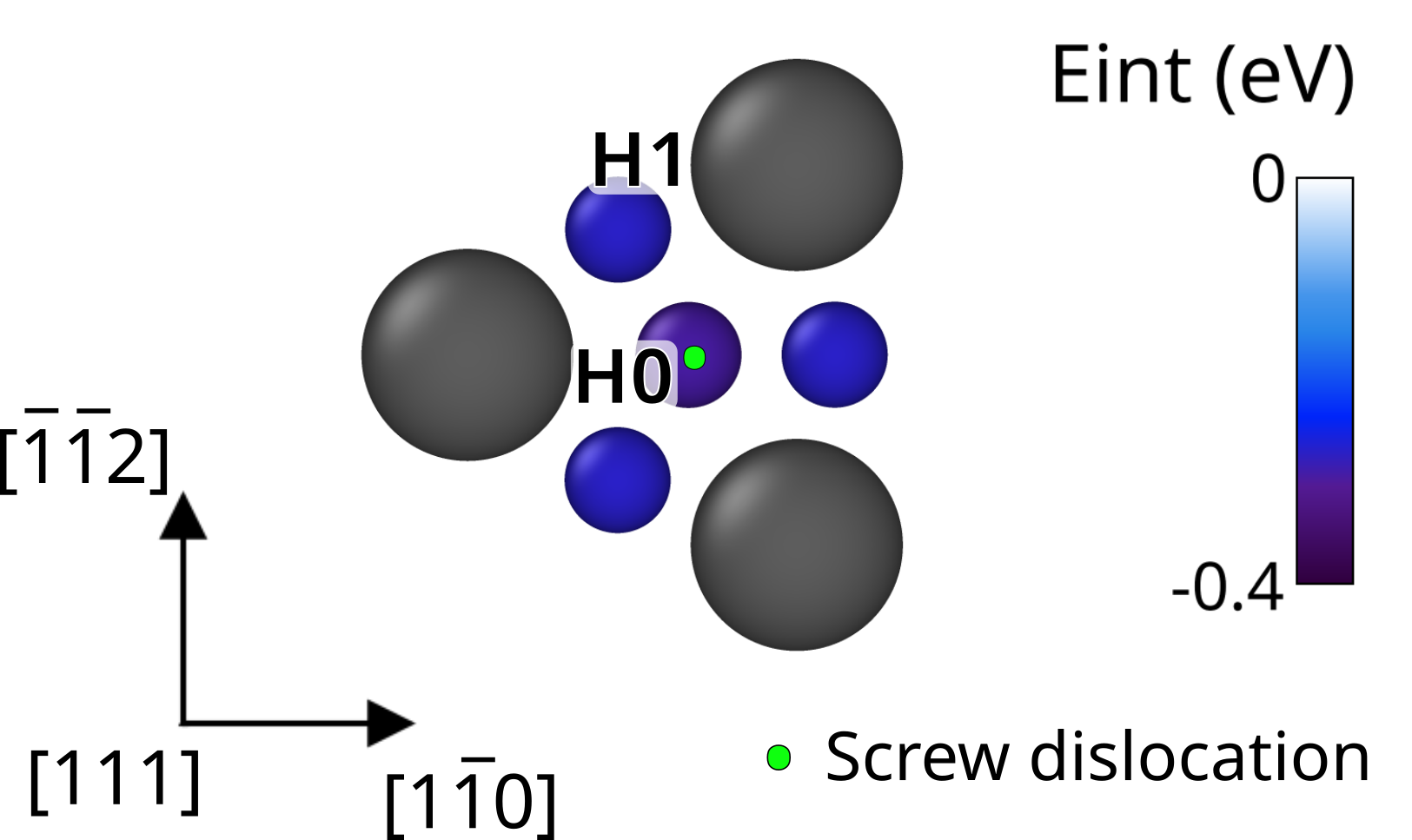}
    \caption{Interaction energy between a hydrogen atom and a screw dislocation in the hard-core configuration at $1h/b$. Iron atoms are shown in grey, and the smaller sphere represents the hydrogen atom in site H0 and H1. The color scale indicates the interaction energy.}
    \label{fig:HC}
\end{figure}

We computed the interaction energy between a hydrogen atom and the screw dislocation in its hard-core configuration, for a high hydrogen concentration along the dislocation line equivalent to 1 H atom per dislocation length of 1$b$. In this case, two preferential trapping sites, usually denoted H0 and H1 (Fig. \ref{fig:HC}), have been identified in the literature \cite{itakura2013effect,simpson2020effect,zheng2023atomistic,huang2023quantitative}. Our results show very good agreement with DFT at comparable concentrations (see Tab. \ref{Tab:HC}).

\begin{table}[h]
\begin{center}
\caption{Interaction energy (eV) between a hydrogen atom and the hard-core configuration at sites H0 and H1, compared with DFT data from the literature \cite{itakura2013effect,simpson2020effect,zheng2023atomistic,huang2023quantitative}.}
\label{Tab:HC}
\begin{tabular}{ l | c | c }
\hline \hline
 & This work & DFT \\ \hline
 H0 & -0.294 & -0.286 \cite{itakura2013effect}, -0.263 \cite{simpson2020effect}, -0.30 \cite{zheng2023atomistic}, -0.29 \cite{huang2023quantitative} \\
 H1 & -0.270 & -0.271 \cite{itakura2013effect}, -0.276 \cite{simpson2020effect}, -0.26 \cite{huang2023quantitative} \\
\hline \hline
\end{tabular}
\end{center}
\end{table}

The H0 site exhibits a slightly lower interaction energy (by about 0.02 eV) than the H1 site, making it more attractive for hydrogen.

It is important to note, however, that for hydrogen concentrations exceeding 1H per 3$b$ along the dislocation line, the hard-core configuration becomes metastable. Under these conditions, all relaxed configurations collapse into the minimum-energy easy-core state.

\begin{figure}[ht!]
    \centering
    \includegraphics[width=1\linewidth]{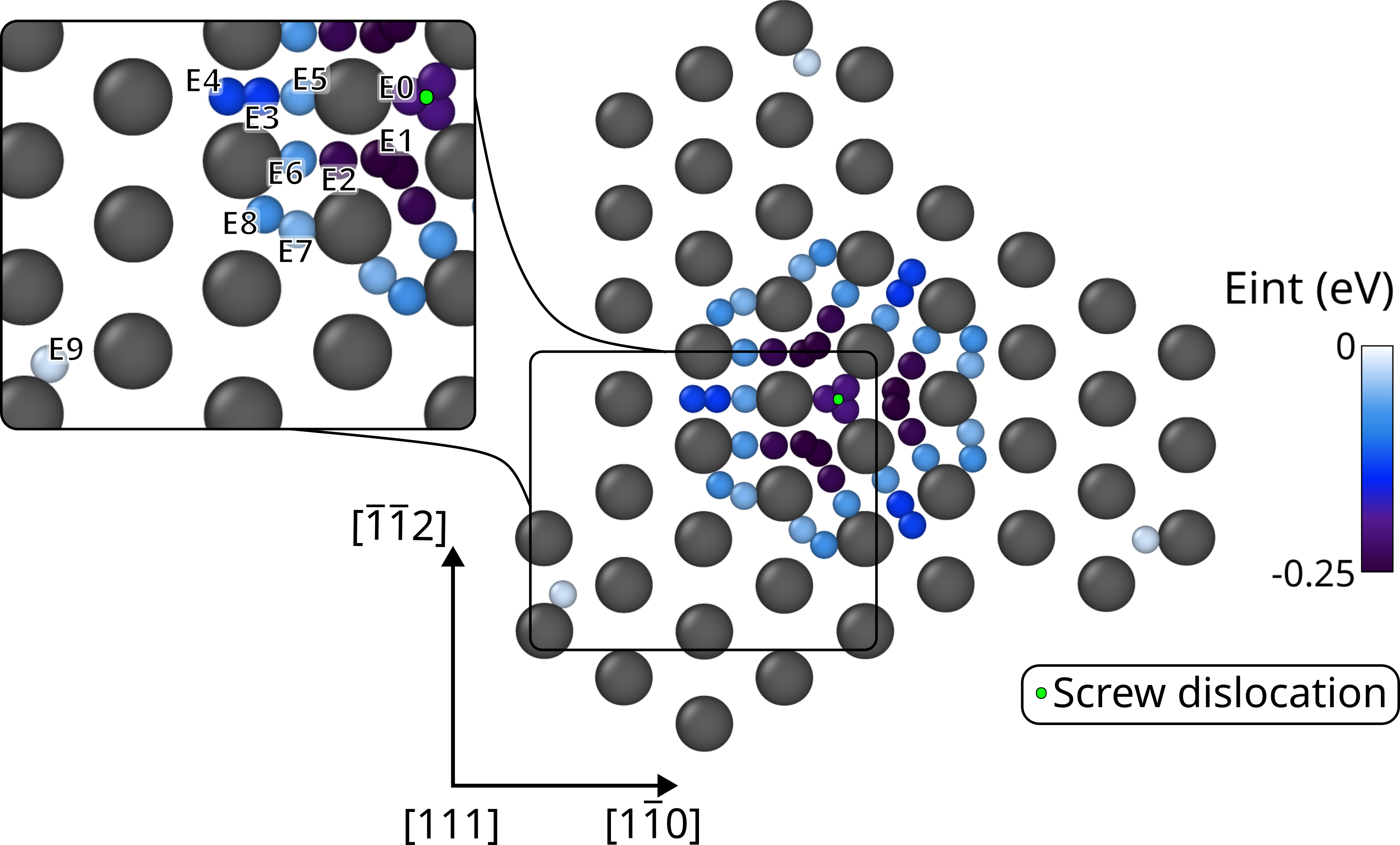}
    \caption{Interaction energy between a hydrogen atom and a screw dislocation in the easy-core configuration. Iron atoms are shown in grey, while the smaller spheres indicate the various hydrogen positions. The color scale represents the corresponding interaction energy.}
    \label{fig:EC}
\end{figure}

We next computed the interaction energies for hydrogen located near the easy-core configuration of the screw dislocation. Ten distinct sites, usually denoted E$i$ in the literature \cite{itakura2013effect,simpson2020effect,zheng2023atomistic,huang2023quantitative} with $i = 0,\ldots,9$, and shown in Fig.~\ref{fig:EC}, were investigated. The corresponding interaction energies are reported in Tab.~\ref{Tab:EC}, together with the DFT reference values. Once again, the potential reproduces the DFT trends with very good accuracy. To enable a direct comparison with DFT, these calculations were initially performed at a high hydrogen concentration (1H/3$b$). However, such strong loading conditions induce significant H–H interactions along the dislocation line, in particular for sites close to the center of the core: E0 to E2 and E5, where non linear effects are present (see Tab. \ref{Tab:EC}).


\begin{table}[h]
\begin{center}
\caption{Interaction energy (eV) between a hydrogen atom and the easy-core configuration at various sites, compared with DFT data from the literature \cite{itakura2013effect,simpson2020effect,zhao2011qm}.}
\label{Tab:EC}
\begin{tabular}{ l | c | c | c }
\hline \hline
Site & This work  &This work  & DFT  \\
  & (1H/10b) & (1H/3b) &  (1H/3b) \\ \hline 

E0 &-0.207 &-0.201 & -0.085 \cite{itakura2013effect}, -0.119 \cite{simpson2020effect} \\ 
E1 &-0.239 &-0.252 & -0.193 \cite{itakura2013effect}, -0.25 \cite{zhao2011qm} \\
E2 &-0.215 &-0.233 & -0.185 \cite{itakura2013effect}, -0.227 \cite{simpson2020effect}, -0.22 \cite{zhao2011qm} \\ 
E3 &-0.127 &-0.137 & -0.138 \cite{itakura2013effect}, -0.134 \cite{simpson2020effect}, -0.15 \cite{zhao2011qm} \\ 
E4 &-0.120 &-0.131 & -0.122 \cite{itakura2013effect}, -0.15 \cite{zhao2011qm} \\
E5 &-0.075 &-0.061 & -0.030 \cite{itakura2013effect} \\
E6 &-0.070 &-0.068 & -0.010 \cite{itakura2013effect} \\
E7 &-0.050 &-0.049 & 0.006 \cite{itakura2013effect} \\ 
E8 &-0.068 &-0.074 & -0.046 \cite{itakura2013effect}, -0.030 \cite{simpson2020effect} \\ 
E9 &-0.018 &-0.018 & -0.005 \cite{itakura2013effect} \\ \hline \hline

\end{tabular}
\end{center}
\end{table}
To isolate the intrinsic interaction between hydrogen and the dislocation core, we subsequently evaluated the interaction energies in the dilute limit, where interactions between hydrogen atoms become negligible. Our simulations indicate that a minimum dislocation-line length of 10$b$ is required to suppress H–H interactions and recover the dilute behavior (see Appendix \ref{App:cellsize}).


The results show that the strongest interactions occur at the core sites E0, E1, and E2, with E1 being the most attractive among them (Tab. \ref{Tab:EC}). The following sites, E3 and E4, exhibit interaction energies nearly a factor of two weaker, and beyond that (sites E5 to E9) the interaction becomes very small. This confirms that the hydrogen–dislocation interaction is predominantly short ranged, as expected for a screw dislocation core.

\begin{figure}[ht!]
    \centering
    \includegraphics[width=1\linewidth]{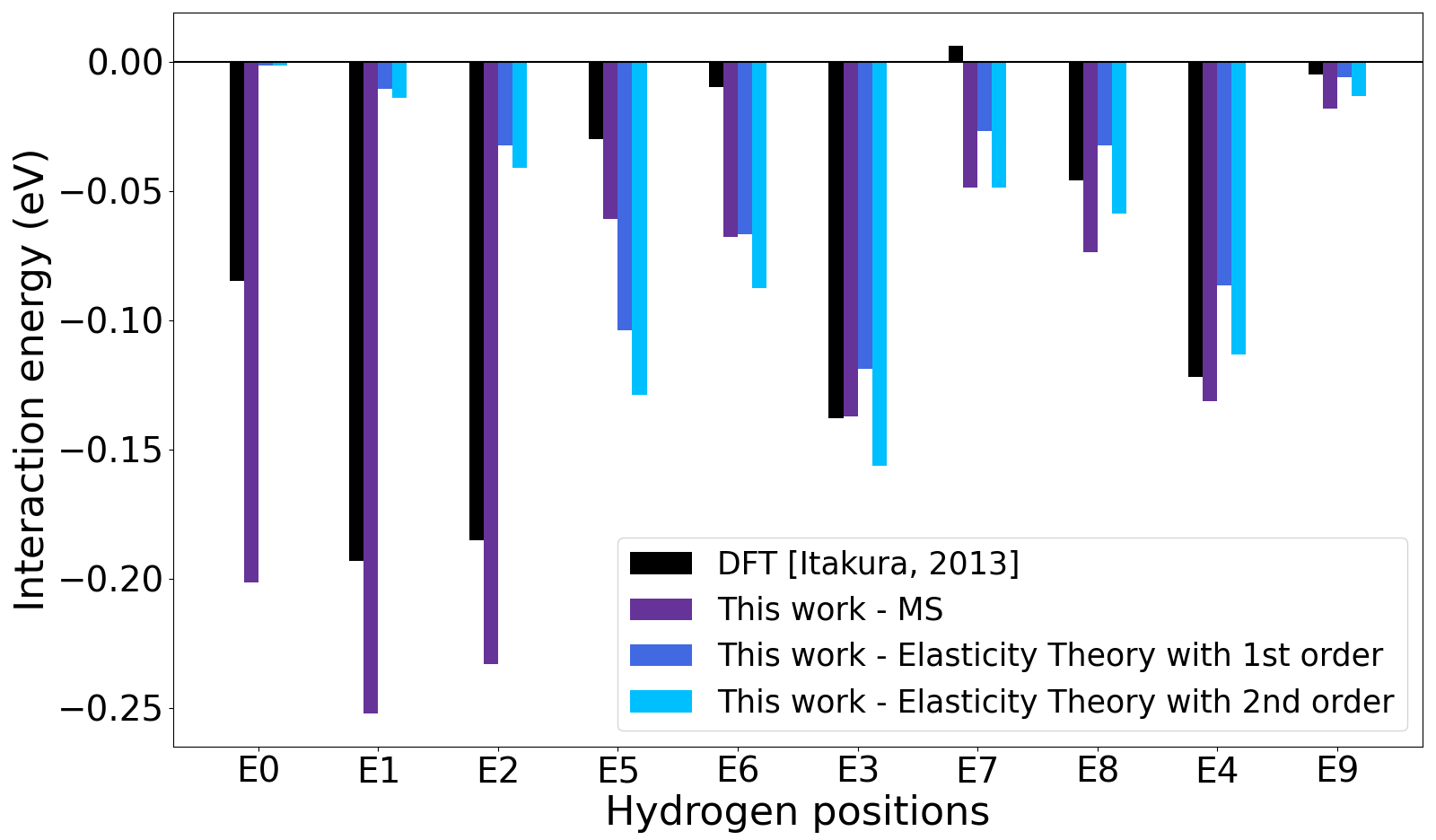}
    \caption{Comparison of the interaction energy in the easy-core configuration obtained from molecular statics simulations, elasticity theory, and DFT calculations. Sites are sorted by distance from the core center.}
    \label{fig:MSvsEL}
\end{figure}

It is worth noting that this interaction energy can be compared with the one predicted by an elasticity theory, using the elastic dipole tensor approach \cite{Gillan_1984, nazarov2016first, clouet2018elastic}. In this framework, we denote the theoretical interaction energy as $E_{int}^{th}$:
\begin{equation}
    E_{int}^{th} = - P^{H}_{ij}\,\epsilon^{D}_{ij},
\end{equation}
where $P^{H}_{ij}$ is the elastic dipole tensor of a hydrogen atom and $\epsilon^{D}_{ij}$ is the strain tensor associated with the screw dislocation.
The strain field $\epsilon^{D}_{ij}$ is directly extracted from our atomistic configurations. 
{
In the second order approximation elastic dipole tensor is expressed as 
\begin{equation}
    P^H_{ij} = P^H_{ij,0} + \alpha_{ijkl}\epsilon^{D}_{kl}
\end{equation}
First order approximation is commonly used, for which $P^H_{ij} \simeq P^H_{ij,0}$ \cite{Gillan_1984, nazarov2016first, clouet2018elastic}.
}
The first order term $P^{H}_{ij,0}$ can be decomposed into a hydrostatic contribution, quoted $P_{ij}^{H,hyd}$, and a deviatoric one, $P_{ij}^{H,dev}$. In the standard crystallographic orientation, this decomposition yields:
\begin{equation}
        P_{ij}^{H,hyd} =
        \begin{pmatrix}
            5.73 & 0 & 0 \\
            0 & 5.73 & 0 \\
            0 & 0 & 5.73
        \end{pmatrix}    
    \label{mat.Pij_hydro}
\end{equation}

\begin{equation}
P_{ij}^{H,dev}=
\begin{pmatrix}
 0 & 0.10 & 0.18 \\
 0.10 & 0.07 & -0.13 \\
 0.18 & -0.13 & -0.08
\end{pmatrix}\label{matricePijdevia}
\end{equation}
Yet the contribution of the deviatoric component is negligible compared with the hydrostatic part (see Appendix \ref{App:pij}).
{The second order term $\alpha_{ijkl}$ provided by the potential is:
\begin{equation}
 \alpha_{ijkl} =
\begin{pmatrix}
\alpha_{11} & \alpha_{12} & \alpha_{12} & 0 & 0 & 0\\
\alpha_{12} & \alpha_{11} & \alpha_{12} &0 & 0 & 0 \\
\alpha_{12} & \alpha_{12} & \alpha_{11} &0 & 0 & 0 \\
0 & 0 & 0 & \alpha_{44} & 0 & 0 \\
0 & 0 & 0 & 0 &\alpha_{44} &  0 \\
0 & 0 & 0 & 0 & 0 & \alpha_{44} 

\end{pmatrix}
\end{equation}
with $\alpha_{11}=7.52$ eV, $\alpha_{12}=9.55$ eV and $\alpha_{44}=-9.08$ eV.
}

The comparison between our molecular statics results and the predictions of elasticity theory is presented in Fig.~\ref{fig:MSvsEL}. { The second-order approximation slightly improves the agreement but remains on the same order of magnitude error as the first-order approximation.} As expected, continuum elasticity fails to capture the complexity of the interaction within the dislocation core region (E0, E1, E2). However, once sufficiently far from the core ($\simeq 1$b), the elastic description becomes relevant and reproduces the interaction energy with good accuracy. 

\subsection{Thermodynamic approach}
We consider a volume of iron $V_{Fe}$ in canonical thermodynamic equilibrium.  
The number of Fe atoms is therefore $n_{Fe} = V_{Fe}\,d_{Fe}$, where $d_{Fe} = 2/a_0^3 = 8.8 \times 10^{28}$at./m$^{3}$ is the atomic density of iron. We denote by $\rho_D$ the dislocation density and by $\xi_H$ the hydrogen solubility in bulk iron. The values used are within the same order of magnitude as those reported in the experimental literature \cite{andrew1950, darken1949, moro2009fragilisation}.

The number of hydrogen atoms in $V_{Fe}$ is then $n_H = \xi_H n_{Fe}$, and the total length of screw dislocation lines within the volume is given by $L = \rho_D V_{Fe}$. In our calculations, we considered that the dislocation line is composed of segments in both easy-core and hard-core configurations. Accordingly, the total dislocation length is written as $ L = (n_{HC} + n_{EC})\, b$, where $n_{HC}$ and $n_{EC}$ denote the number of line elements in the hard-core and easy-core configurations, respectively.

Restricting our analysis to screw dislocations, we determine at thermodynamic equilibrium the probability that a hydrogen atom occupies a specific binding site in the vicinity of the dislocation. The probability of the system being in microstate, corresponding to a hydrogen atom occupying site type $S_i \in$ (H0, H1, E0,E1,E2,...,E8, bulk tetrahedric and octaedric sites), is:

\begin{equation}
    p_{S_i} = 
    \frac{g(E_{S_i}) \exp\!\left(-\beta E_{S_i}\right)}{Z},
    \label{eq.proba}
\end{equation}
where $E_{S_i}$ is the interaction energy between a screw dislocation and a hydrogen atom, extracted directly from our molecular statics simulations (Eq. \ref{eq:Eint}), and $g(E_{S_i})$ denotes the degeneracy of site type $S_i$, arising from the symmetries of the system. The corresponding partition function $Z$ is
\begin{equation}
    Z = \sum_{E_{S_i}}^{N} g(E_{S_i}) \exp\!\left(-\beta E_{S_i}\right).
\end{equation}
In the present work, the summation includes all dislocation–core sites previously characterized, namely sites E0 to E8 in the easy-core configuration and sites H0 and H1 in the hard-core configuration. All remaining accessible positions available to hydrogen are taken as regular interstitial sites in the bcc lattice, i.e., the tetrahedral sites (3$n_{Fe}$) and octahedral sites (6$n_{Fe}$).

This framework allows us to assess whether the core sites of the screw dislocation can become saturated, as often assumed in the literature. It should be emphasized, however, that this approach neglects other microstructural defects, such as edge dislocations, grain boundaries, and vacancies, that may act as deeper and more numerous traps than screw dislocations. As a consequence, the saturation level obtained here represents an upper-bound estimate.

Within this framework, the saturation ratio $\eta_{S_i}$ of the $S_i$ core sites is defined as the fraction of available sites that are statistically occupied:

\begin{equation}
    \eta_{S_i} = 
    \frac{p_{S_i}\, n_H}{g(E_{S_i})\, L}.
    \label{eq.ratio}
\end{equation}

A site type $S_i$ is considered saturated when  $g(E_{S_i}) L\leq p_{S_i}\, n_H.$

\begin{figure}[ht!]
    \centering
    \includegraphics[width=0.7\linewidth]{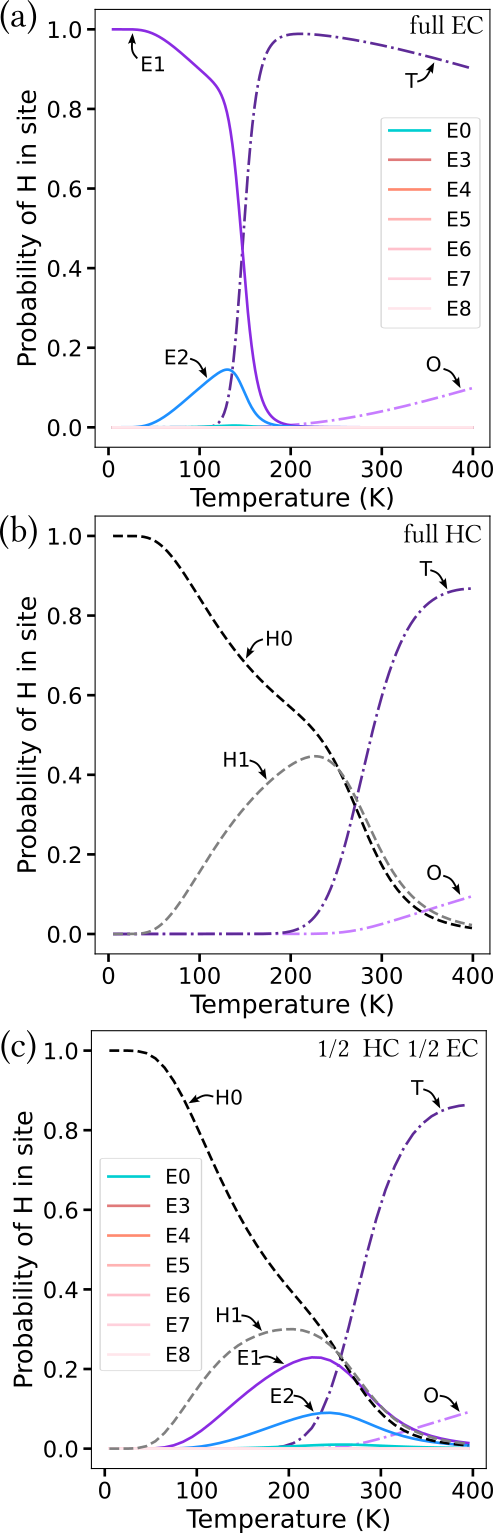}
    \caption{Probability $p_{S_i}$ for a hydrogen atom to occupy site $S_i$, obtained from Eq. \ref{eq.proba}, for $\xi_H = 10^{-5}$ at. and $\rho_D = 10^{15}$ m$^{-2}$, in the cases of (a) a fully easy core, (b) a fully hard core, and (c) a mixed configuration composed of half easy core and half hard core. Bulk sites are denoted T for tetrahedral sites and O for octahedral sites.}
    \label{fig:proba}
\end{figure}

Fig. \ref{fig:proba} presents the results obtained from Eq. \ref{eq.proba}, $i.e$ the probability for hydrogen to occupy site ${S_i}$, for a fixed solubility $\xi_H = 10^{-5}$ at. and a dislocation density $\rho_D = 10^{15}$ m$^{-2}$, as a function of temperature. Three configurations were considered: (a) a fully easy core, (b) a fully hard core, and (c) a mixed configuration composed of half easy core and half hard core.
At low temperature, only the sites exhibiting the strongest interaction with the dislocation core show a significant occupation probability: E1 and E2 in case (a), H0 and H1 in case (b), and H0, H1, E1, and E2 in case (c). However, as the temperature increases, reaching values close to room temperature, hydrogen preferentially occupies bulk tetrahedral lattice sites due to their high degeneracy (see Fig. \ref{fig:proba}).

\begin{figure}[ht!]
    \centering
   \includegraphics[width=0.9\linewidth]{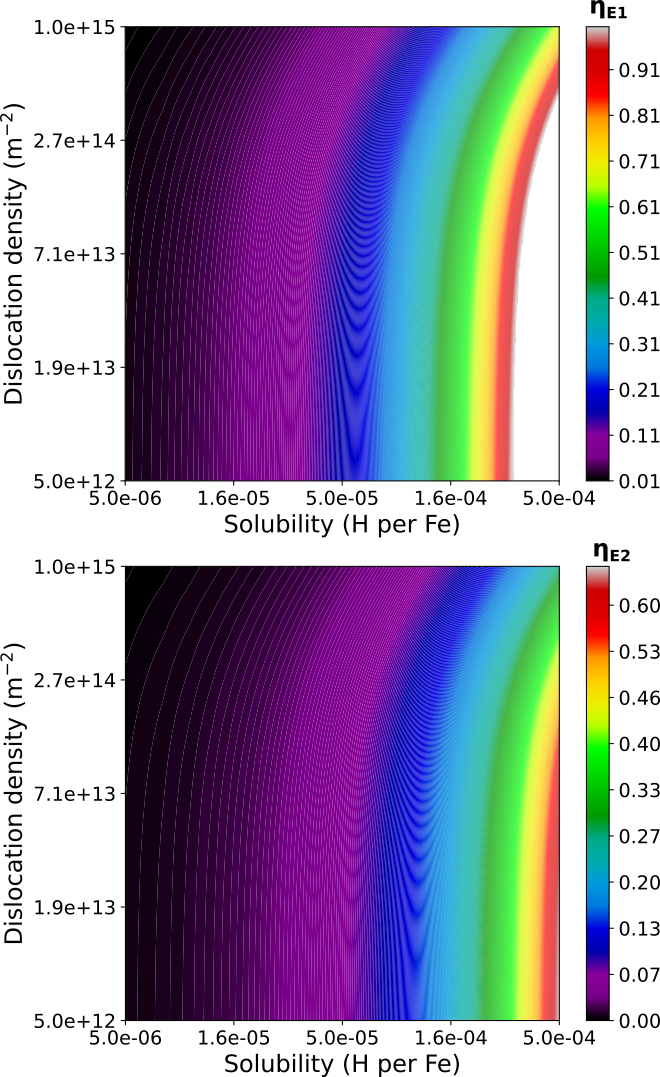}
    \caption{Saturation ratio of the $E1$ and $E2$ sites as a function of the dislocation density $\rho_D$ and the hydrogen solubility $\xi_H$, for $L=n_{EC}b$. White regions indicate the range of ($\rho_D,\xi_H$) leading to full saturation of the sites by hydrogen.}
    \label{fig:fullEC}
\end{figure}

\begin{figure}[ht!]
    \centering
    \includegraphics[width=0.9\linewidth]{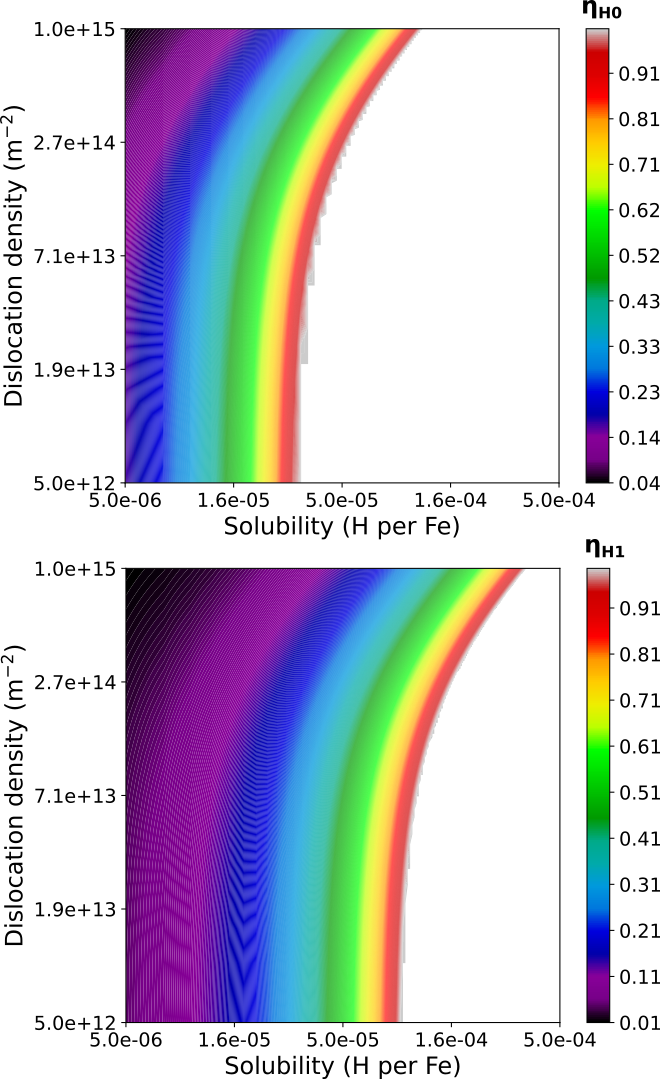}
    \caption{Saturation ratio of the $H0$ and $H1$ sites as a function of the dislocation density $\rho_D$ and the hydrogen solubility $\xi_H$, for $L=n_{HC}b$.  White regions indicate the range of ($\rho_D,\xi_H$) leading to full saturation of the sites by hydrogen.}
    \label{fig:fullHC}
\end{figure}

We then computed $\eta_{S_i}$, the saturation ratio at room temperature ($T = 300$ K), for such most probable sites (E1, E2, H0, and H1), according to Eq. \ref{eq.ratio}.

First we considered two limiting cases: (i) a configuration in which the entire dislocation line is assumed to adopt the easy-core structure, for which $L = n_{EC} b$ (see Fig.~\ref{fig:fullEC}); and (ii) a configuration where the whole line is taken to be in the hard-core structure, giving $L = n_{HC} b$ (see Fig.~\ref{fig:fullHC}).

It is worth noting that, in the full easy-core case (Fig.~\ref{fig:fullEC}), the saturation ratio remains close to 0.2, meaning that only about one fifth of the dislocation line is occupied, within the range of hydrogen solubilities and dislocation densities (linked to the applied strain rate) that is experimentally accessible (see App. \ref{App:dislo_density}, Fig.~\ref{fig:exphydsolu}). However, such a level of saturation is likely overestimated, since real materials contain numerous other deep trapping sites (edge dislocations, grain boundaries, vacancies, etc.) that compete with screw dislocations for hydrogen. 

Our results contribute to the ongoing discussion regarding the hydrogen saturation of screw dislocations in bcc iron by including easy-core sites and dislocation density. In particular, they suggest that the commonly assumed full saturation of the screw-dislocation core is, in practice, unlikely to occur in bcc Fe under gaseous hydrogen loading.
\begin{figure}[ht!]
    \centering
    \includegraphics[width=0.8\linewidth]{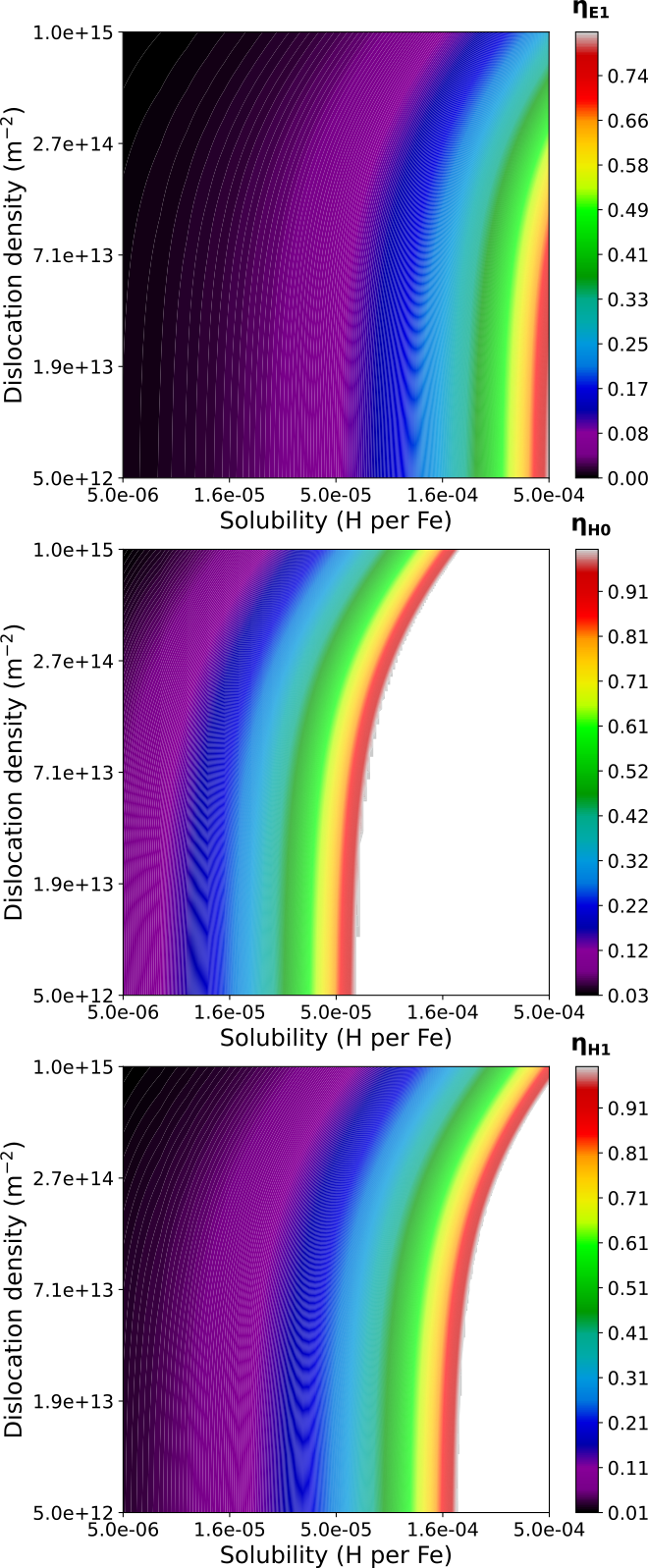}
    \caption{Saturation ratio of the E0, H0, and H1 sites as a function of the dislocation density $\rho_D$ and the hydrogen solubility $\xi_H$, for $L = (n_{HC} + n_{EC})b$, where half of the dislocation line adopts the easy-core configuration and the other half the hard-core configuration. White regions indicate the range of $(\rho_D,\xi_H)$ values leading to full saturation of these sites by hydrogen.}
    \label{fig:ECHC}
\end{figure}

Indeed, when assuming that the entire dislocation adopts a hard-core configuration, the core becomes fully saturated with hydrogen. However, such a scenario is energetically unlikely. The hard-core structure is intrinsically metastable and is only stabilized under conditions of extreme hydrogen supersaturation. This leads to a self-consistent but unrealistic situation: the hard-core configuration would require a high hydrogen concentration to exist, yet its apparent strong trapping capacity arises precisely because of that artificially imposed supersaturation in sites H0 and H1. In other words, the hard-core configuration appears saturated largely because the conditions required for its stability already enforce an unrealistically high hydrogen content.

Fig.~\ref{fig:ECHC} illustrates a configuration in which half of the dislocation line adopts the hard-core structure and the other half the easy-core structure. This situation is more realistic than the fully hard-core case, although it still represents a strongly hydrogen-loaded state.

In this third configuration, our results show that more than half of the $H_1$ sites become occupied under experimentally accessible values of $\rho_D$ and $\xi_H$ (see App. \ref{App:dislo_density}). In contrast, the easy-core sites exhibit a much lower probability of saturation, and only about one fifth of the H0 sites reach full occupancy.

\section{\label{Conclusion}Conclusion}
For both the hard-core and easy-core configurations, the interaction energies obtained with the NNIP potential closely reproduce available DFT data. In the hard-core configuration, two preferential binding sites (H0 and H1) are identified. However, this configuration is metastable and only stabilized under conditions of strong hydrogen supersaturation; at lower concentrations, dislocation spontaneously relaxes toward the easy-core state.

The easy-core configuration exhibits several binding sites, among which E0, E1, and E2 show the strongest interactions, with E1 being the most attractive. The interaction rapidly decreases away from the core, confirming its short-range character. Comparison with continuum elasticity further indicates that linear elastic predictions are reliable only beyond the core region, whereas atomistic effects dominate within the core. This clear delineation of the validity range of elastic modeling will be very useful for future large-scale simulations.

The thermodynamic analysis reveals that, even in the limiting case where the entire dislocation line adopts the easy-core configuration, the saturation ratio remains modest: only about one fifth of the core sites become occupied under experimentally accessible hydrogen solubilities and dislocation densities. This estimated saturation level must be considered an upper bound, since other microstructural defects, such as edge dislocations, vacancies, and grain boundaries, compete for hydrogen and act as deeper traps than screw dislocations.

Assuming a fully hard-core dislocation line implies complete saturation of the H0 and H1 sites. However, this scenario appears unrealistic for bcc Fe under gaseous hydrogen loading, given the metastability of the hard-core configuration and its dependence on extreme hydrogen supersaturation. A mixed configuration, with half the line in easy core and half in hard core, yields intermediate behavior: H1 sites become significantly filled, while easy-core sites remain only weakly occupied.

Altogether, these results challenge the frequently invoked assumption that the screw-dislocation core in bcc iron becomes fully saturated with hydrogen under service conditions. Instead, our findings demonstrate that such saturation is unlikely, and that hydrogen occupancy of screw-dislocation cores should remain partial in realistic microstructural and environmental conditions.

\section{\label{Appendix}Appendix}

\subsection{Cell size effect \label{App:cellsize}}

\begin{figure}[ht!]
    \centering
    \includegraphics[width=8.5cm]{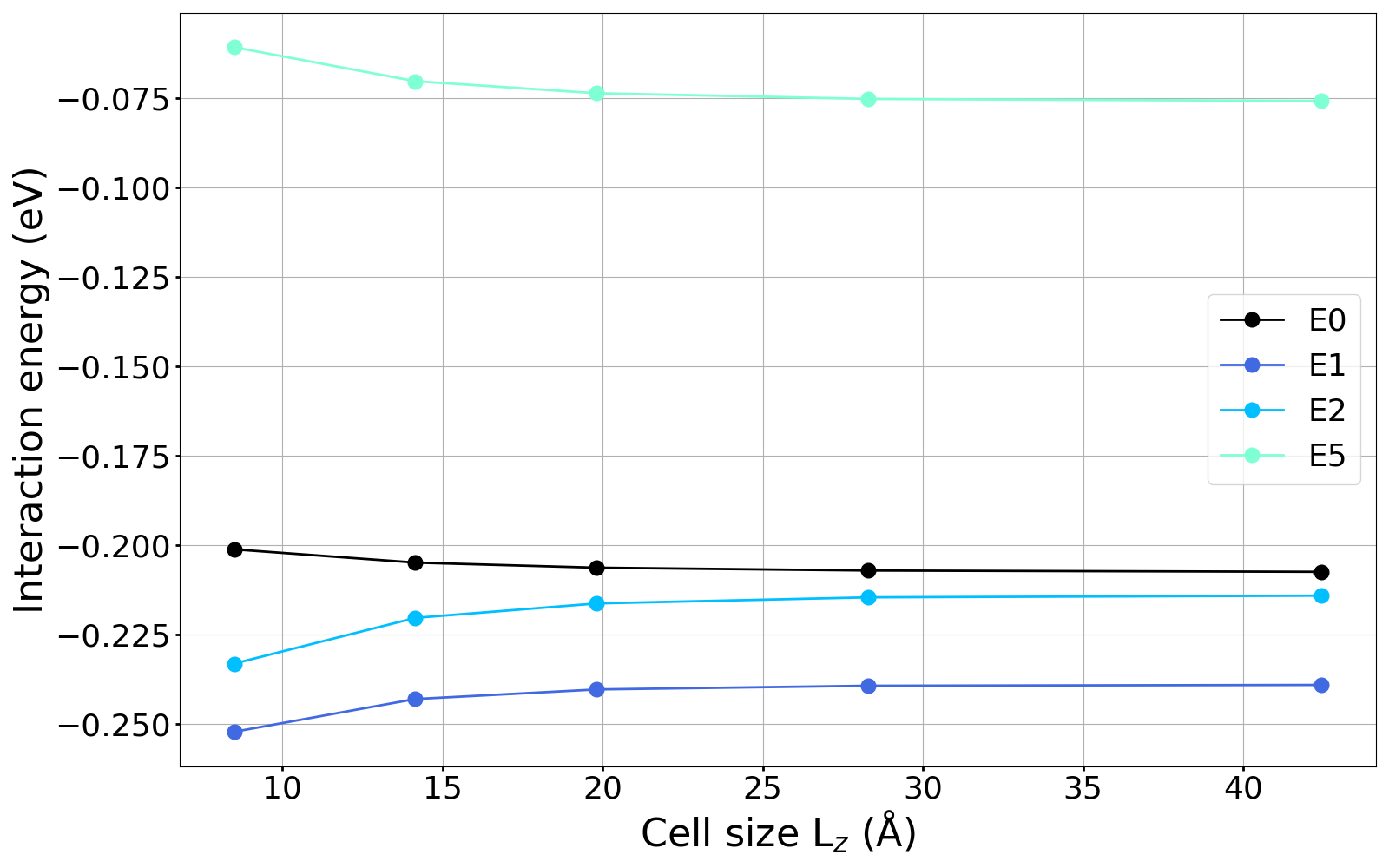}
    \caption{Interaction energy between a hydrogen atom, at the core site (E0, E1, E2 and E5) respectively, and an easy core dislocation, according to the cell size along $z$-direction. }
    \label{fig:cell_size}
\end{figure}

Fig. \ref{fig:cell_size} shows the evolution of interaction energy between a hydrogen atom and a dislocation with the cell size along the $z$-direction, which corresponds to the dislocation line length. The interaction energy is defined according to Eq. \ref{eq:Eint}. Computations are shown for a hydrogen in the easy core dislocation, at site E0, E1 and E2, respectively, but similar results were observed for the other sites.
For these sites, the convergence of the interaction energy is reach above $L_z$ = 28.3 \AA.
Below this dislocation line length, the hydrogen atom interacts with its periodic images. It is worth noting that this interaction is influenced by the elastic strain response of the dislocation, as the hydrogen interaction depends on the specific site. E0 and E5 sites exhibit a long-range repulsive interaction, in contrast to the E1 and E2 sites. 


\subsection{Decomposition of $P_{ij}$ \label{App:pij}}
\begin{figure}[ht!]
    \centering
    \includegraphics[width=8.5cm]{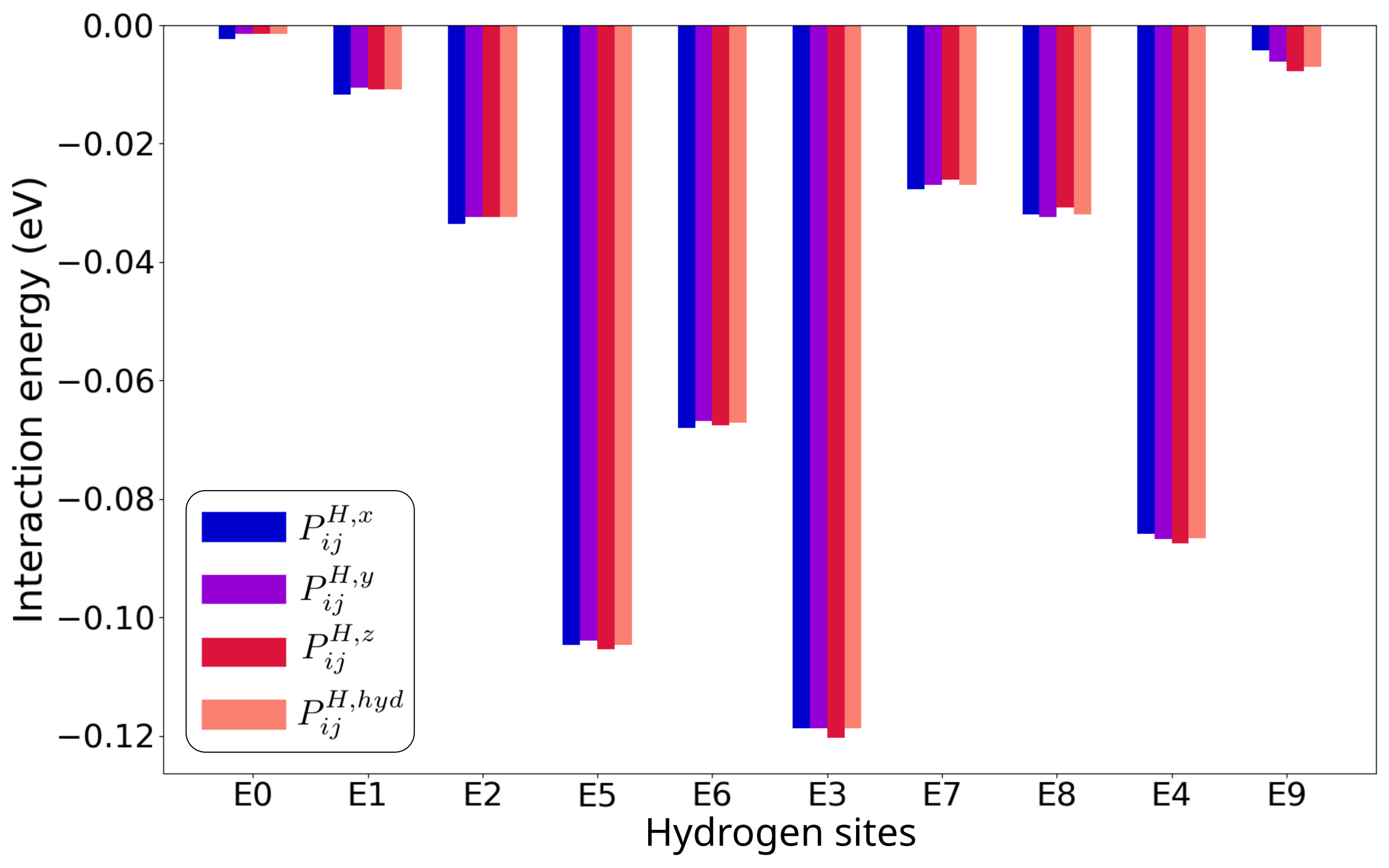}
    \caption{Interaction energy computed within the frame of the elasticity theory (Eq. \ref{eq:Eint}), using several expression of the elastic dipole tensor (Eqs. \ref{mat.Pij_hydro},\ref{eq:dipz}-\ref{eq:dipx}).}
    \label{fig:DiagPij+}
\end{figure}

Due to the structure and symmetries of the octahedral site, the elastic dipole tensor $P^H_{ij,0}$ is, in principle, anisotropic and thus its expression depends on the hydrogen position within the iron lattice. Depending on the specific site, the elastic dipole tensor can take one of the following diagonal forms, for the different octahedral sites:
\begin{equation}
    P^{H,z}_{ij} = 
    \begin{pmatrix}
        p_{1} & 0 & 0 \\
       0 & p_{1}  & 0 \\
        0 & 0 & p_{2} \\
    \end{pmatrix}
    \label{eq:dipz}
\end{equation}
or 
\begin{equation}
    P^{H,y}_{ij} = 
    \begin{pmatrix}
        p_{1} & 0 & 0 \\
       0 & p_{2}  & 0 \\
        0 & 0 & p_{1} \\
    \end{pmatrix}
    \label{eq:dipy}
\end{equation}
or
\begin{equation}
    P^{H,x}_{ij} = 
    \begin{pmatrix}
        p_{2} & 0 & 0 \\
       0 & p_{1}  & 0 \\
        0 & 0 & p_{1} \\
    \end{pmatrix}
    \label{eq:dipx}
\end{equation}
with $p_1= 5.87$ eV and $p_2= 5.44$ eV. 
This does not affect the energy under isotropic strain conditions but may induce energy variations near the dislocation core. 
Fig. \ref{fig:DiagPij+} shows the difference between $P^{H,x}_{ij}$, $P^{H,y}_{ij}$, $P^{H,z}_{ij}$, and $P^{H,hyd}_{ij}$. No significant energy variation is observed. Hence, we may reasonably assumed that $P^{H}_{ij,0} \simeq P^{H,hyd}_{ij}$, which simplifies the expression to its hydrostatic form.


\subsection{H solubility as function of dislocation density \label{App:dislo_density}}

\begin{figure}[ht!]
    \centering
    \includegraphics[width=8cm]{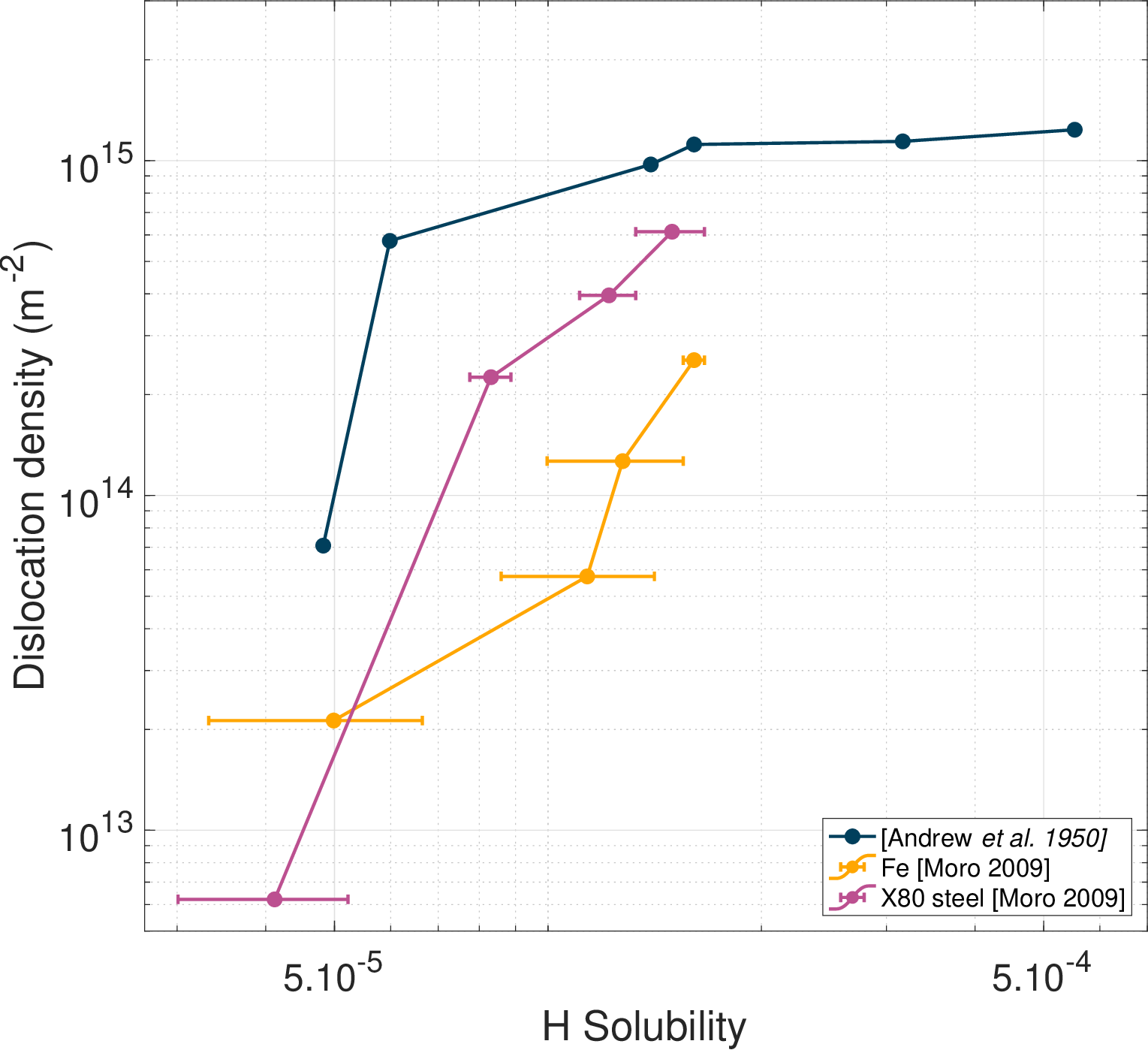}
    \caption{Correlation between H solubility measure experimentally and dislocation density in low C steels. Dislocation densities are estimated from the mechanical properties.}
    \label{fig:exphydsolu}
\end{figure}


Hydrogen solubility was reported from experiments in pure Iron and low carbon steels in \cite{andrew1950, darken1949, moro2009fragilisation} conducted at room temperature for different amount of strain $\epsilon$. Total dislocation density may be estimated from hardness measures \cite{andrew1950} or macroscopic deformation curves $\sigma(\epsilon)$ \cite{moro2009fragilisation}, when provided. The polycrystalline flow stress $\sigma$ is typically related to the mesoscopic critical shear stress on active slip systems $\tau_c$ as:
\begin{equation}
    \sigma = M(\epsilon) \tau_c  
\end{equation}

where $M$ is the so-called Taylor coefficient. Hardening H0 typically scales as H0 $\propto C \sigma$ with $C\approx3$ the so-called Tabor's coefficient \cite{tabor2000}. Initially, $M$ starts $\approx 2.75$ for random crystalline orientation and when considering only ${a_0/2}<111> \left\{110 \right\}$ as the main slip systems \cite{queyreau2009}. We chose a linear increase of $M(\epsilon)$ as deformation increases up to  $\approx 3.2$ for large deformation.

The critical shear stress of bcc systems is commonly modeled at the mesoscale as \cite{queyreau2009,kahloun2013,monnet2013,Queyreau:2024}:
\begin{equation}
    \tau_c = \tau_{P} + \tau_{SS} + \alpha(\rho) \mu b \sqrt{\rho} \; \left(+\mu b /{w_L}\right)
    \label{eq:tau_c}
\end{equation}
where $\tau_{P}$ is the friction stress associated to the thermally activated motion of screw dislocations \cite{kahloun2013,monnet2013}. The friction stress, while small of the order of $\tau_{P}\approx$ 7 MPa, is still noticeable at room temperature in iron single crystals \cite{takeuchi1975}. The second term $\tau_{SS}$ in the rhs represents the solid solution contribution to the overall strengthening. C interstitial atoms contributes the most to this term in Fe and ferritic steels. Solid solution strengthening typically scales with the the carbon content $c$ as $\tau_{SS}\propto \sqrt{c}$ \cite{queyreau2009}. For the low carbon steels considered here, C in solid solution corresponds to $ c\approx 60-75$ ppm, and $\tau_{SS}$ can be estimated of about $\tau_{SS}\approx 20-27$ MPa. When present, Si and Mn contents may contribute to $\tau_{SS}$ as well as $(\textrm{Ni wt}\%) \times83.2+(\textrm{Mn wt} \%) \times 32.3$ in MPa \cite{pickering1978}. The last contribution in Eq. \ref{eq:tau_c}, is the so-called forest strengthening induced by the dislocation density $\rho$. The scalar interaction coefficient $\alpha$ averages the various interactions among $\left\{110 \right\}$ systems, which are weaker than the fcc counterparts, yielding to $\alpha\approx 0.3$ at $\rho = 10^{12}$ m$^{-2}$ \cite{queyreau2009}. $\alpha(\rho)$ is known to decrease with dislocation density $\rho$ due to the logarithmic dependence of the dislocation self stress \cite{queyreau2009,Queyreau:2024}. Assuming that all stress contributions are constant at the exception of forest hardening, we solved Eq. \ref{eq:tau_c} iteratively to estimate $\rho$.

The results from this analysis are understandably scattered. The estimated dislocation densities are shown in Fig. \ref{fig:exphydsolu} as a function of the measured hydrogen solubility. As expected for bcc Fe, the hydrogen solubility in undeformed samples is rather low, about $5\times10^{-5}$ H per Fe atom ($\simeq$ 1 wppm). Hydrogen solubility then increases in good correlation with dislocation density up to approximately $5 -7\times10^{14}$ m$^{-2}$, with dislocations arguably acting as the main trapping sites for H atoms. Beyond this range, the dislocation density tends to saturate around a few $10^{15}$ m$^{-2}$, comparable to the saturation values observed in deformed single crystals of pure Fe \cite{spitzig1970,takeuchi1975}. At higher strains, although the dislocation density saturates, the hydrogen solubility continues to increase with deformation without showing a clear slowdown, suggesting that additional hydrogen trapping mechanisms must be involved. For heavily deformed ferritic steels ($\epsilon > 0.5$) \cite{andrew1950, darken1949}, the maximum H solubility was found close to $5\times10^{-4}$ H per Fe atom ($\simeq$ 10 wppm).

\section{\label{Acknowledments}Acknowledgments}
The computational resources utilized for this study were graciously provided by the MCIA (Mésocentre de Calcul Intensif Aquitain) and the NOETHER computing facilities at La Rochelle University, hosted at the LaSIE laboratory and administered by Dr Antoine Falaize, to whom we are particularly grateful. The experiments in this work were carried out on MAGI, the experimental platform of University of Sorbonne Paris North (USPN) dedicated to research. This platform offers researchers at the institution High-Performance Computing (HPC), cloud and storage services administered by Nicolas Greneche.
The authors acknowledge support from the French National Research Agency (ANR) under France 2030 program and reference ANR-PEHY-00 04 (project PEPRH2- Hyperstock).


\bibliography{apssamp}


\end{document}